\documentclass{article}

\usepackage[final]{neurips_2023}

\usepackage[utf8]{inputenc} 
\usepackage[T1]{fontenc}    
\usepackage{hyperref}       
\usepackage{url}            
\usepackage{booktabs}       
\usepackage{amsfonts}       
\usepackage{nicefrac}       
\usepackage{microtype}      
\usepackage{xcolor}         
\usepackage{graphicx}
\usepackage{multirow}
\usepackage{booktabs}
\usepackage{natbib}
\usepackage{array}
\usepackage{colortbl}
\usepackage{booktabs}
\usepackage{microtype}      
\usepackage{xcolor}         
\usepackage{graphicx}
\usepackage{algpseudocode}
\usepackage{algorithm}
\usepackage{amsmath}
\usepackage{hyperref}

\definecolor{darkgray}{gray}{0.7}
\definecolor{lightgray}{gray}{0.9}

\title{Autoregressive fragment-based diffusion for pocket-aware ligand design}

\author{%
  Mahdi Ghorbani \\
  University of California, San Francisco \\
  \texttt{ghorbani@keiserlab.org} \\
  \And
  Leo Gendelev \\
  Genentech \\
  \texttt{gendelev.leo@gene.com } \\
  \And
  Paul Beroza \\
  Genentech \\
  \texttt{berozap@gene.com } \\
  \And
  Michael J. Keiser \\
  University of California, San Francisco \\
  \texttt{keiser@keiserlab.org} \\
}

\begin{document}

\maketitle

\begin{abstract}
    In this work, we introduce AutoFragDiff, a fragment-based autoregressive diffusion model for generating 3D molecular structures conditioned on target protein structures. We employ geometric vector perceptrons to predict atom types and spatial coordinates of new molecular fragments conditioned on molecular scaffolds and protein pockets. Our approach improves the local geometry of the resulting 3D molecules while maintaining high predicted binding affinity to protein targets. The model can also perform scaffold extension from user-provided starting molecular scaffold.
\end{abstract}

\section{Introduction}

Rational drug design against defined binding pockets relies heavily on computational modeling.\citet{stokes2020deep, anderson2003process} Traditionally, the diversity of small-molecule candidates and the high degrees of freedom inherent in ligand-protein binding systems make navigating chemical space computationally intensive.\citet{lipinski2004navigating} Moreover, target-aware molecular design strives to balance optimizing for potency against specific target structures while maintaining desirable absorption, distribution, metabolism, and excretion (ADME) and pharmacokinetic and pharmacodynamic (PKPD) properties.\citet{skalic2019target} While many target protein structures are available, effectively harnessing this information to design novel drug-like compounds with desired therapeutic effects remains an active area of research.\citet{ragoza2017protein} 

Diffusion models \citet{ho2020denoising, kingma2021variational} generate 3D molecular structures from underlying distributions of molecular data\citet{hoogeboom2022equivariant}, thus enabling the generation of diverse molecular candidates that reflect real chemical space. However, these models struggle to capture the nuances of local molecular geometry. Specifically, maintaining the correct spatial arrangements and conformations of functional groups and atoms remains challenging.\citet{harris2023benchmarking} While the overall structure might resemble known molecules, minor deviations in local geometry can significantly impact the bioactivity and specificity of the generated compounds.

Many pocket-specific molecule generation models have leveraged autoregressive strategies. In these models, atoms are placed individually, and bonds are determined separately. \citet{drotar2021structure, liu2022generating}. However, this sequential approach can be cumbersome and error-prone; even generating a benzene ring is a laborious six-step procedure. Fragment-based generation strategies sidestep some of these drawbacks. Our work employs Autoregressive Diffusion Models (ARDMs),\citet{hoogeboom2021autoregressive} which can generate data in a flexible order. This unique feature enables ARDMs to bridge the gap between order-agnostic autoregressive and diffusion-based generative models.\citet{igashov2022equivariant, hoogeboom2022equivariant, schneuing2022structure, guan20233d} Unlike their traditional counterparts, ARDMs don't adhere to strict architectural norms for neural networks, yet they achieve comparable results in fewer steps.

In this study, we combine fragment-based drug design with autoregressive diffusion models. Unlike traditional autoregressive methods that work atom by atom, this combined approach allows each fragment to undergo a denoising process, predicting atom coordinates and atom types (Figure 1). Rather than relying on a fixed fragment library, our approach dynamically generates fragments, providing flexibility in the diversity of fragments produced. This approach generates molecules with more accurate local geometries for pocket-based molecule generation, delivering greater precision and efficiency in drug design.

\begin{figure*}
\centering
  \includegraphics[scale=0.35]{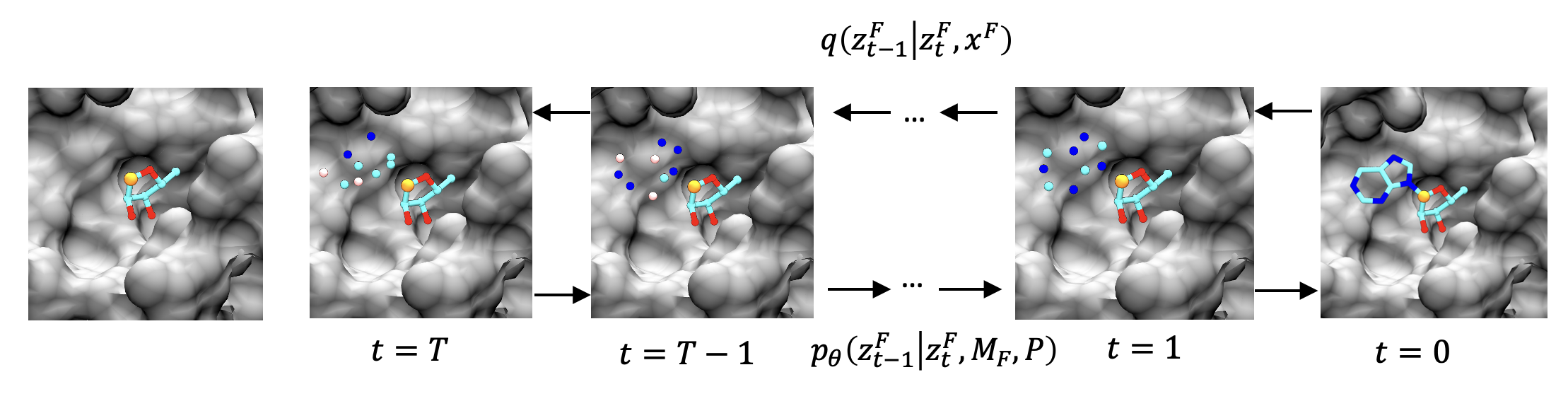}
  \caption{Noising and Sampling for a single fragment inside a protein pocket. Yellow spheres show the anchor point.}
\end{figure*}
\vspace{-0.1cm}

\section{Related Work}
Generative models and geometric deep learning have influenced recent pocket-based drug design. \citet{atz2021geometric, bronstein2017geometric}. \citet{li2021structure} introduced an autoregressive generative model designed to sample ligands, using the pocket as a conditioning constraint. Building on this work, Peng et al. \citet{peng2022pocket2mol} introduced Pocket2Mol, which uses an E(3) equivariant graph neural network\citet{satorras2021n} that accounts for rotation and translation symmetries in 3D space for more accurate molecular representations. Similarly, \citet{drotar2021structure}, \citet{liu2022generating} explored autoregressive models for molecular generation, generating atoms sequentially. These models incorporate angles during generation to improve molecular detail and accuracy. 

Diffusion models enable pocket-free and pocket-based drug design.\citet{kingma2021variational} \citet{hoogeboom2022equivariant} introduced Equivariant Diffusion Models (EDMs), which simultaneously learn continuous coordinates and atom types for molecule generation. Multiple studies built on this approach: GeoDiff \citet{xu2022geodiff} predicts a molecule's 3D conformation and DiffLinker\citet{igashov2022equivariant} learns to connect seed fragments. Similarly, Schneuing et al. \citet{schneuing2022structure} developed DiffSBDD, a denoising diffusion model for pocket-based molecule design. Guan et al's TargetDiff uses SE(3)-equivariant networks to explicitly learn the generative process for continuous coordinates and categorical atom types.\citet{guan20233d}. Peng et al introduced FragDiff \cite{peng2022pocket}, an autoregressive diffusion model on molecular fragments. By comparison, our approach employs order-agnostic autoregressive diffusion models \cite{hoogeboom2021autoregressive}, and its high molecule validity from Geometric Vector Perceptrons \cite{jing2020learning} eliminates the need for a discriminator.

\vspace{-0.2cm}
\section{Methods}
\vspace{-0.2cm}
\subsection{Problem Definition}
We represent the protein pocket and the ligand as point clouds with atomic coordinates $r$ and corresponding feature vectors $h$. The feature vector is the one-hot encoded atom type for ligand atoms and element type, plus amino acid type for the pocket atoms. For the pocket $P={(r^P _i, h^P _i})_{i=1} ^ {N_P}$, and for the molecule $M={(r^M _i, h^M _i})_{i=1} ^ {N_M}$ where $N_P$ and $N_M$ are the number of atoms in the pocket and molecule respectively. We further separate each molecule into multiple fragments and molecular scaffolds $M={[(r^{M_F} _i, r^{F} _i), (h^{M_F} _i, h^{F}_i)}]_{i=1}^ {N_M}$. $M_F$ and $F$ superscripts represent molecule scaffold and the fragment respectively.  Note that for each molecule, there exist multiple fragments and scaffolds. The autoregressive diffusion process aims to generate a new fragment conditioned on a molecular scaffold and protein pocket at each step.

\subsection{Diffusion Process}
The diffusion process iteratively adds noise to data point $x$ and trains a neural network to remove noise progressively (Figure 1). Generative denoising inverts the trajectory when $x$ is unknown. This process for fragment $F$ is conditioned on the molecular scaffold $M_F$ and the protein pocket $P$:
\begin{equation}
    p\left(z_{t-1}^{F} | z_t^{F}, M_F, P\right) = q\left(z_{t-1}^{F} | \hat{x}^{F}, z_t^{F}\right)
\end{equation}
where $\hat{x} = (1/\alpha_t) z_t - (\sigma_t / \alpha_t)\hat{\epsilon_t}$ is the approximation of $x^F$ computed by neural network $\phi$ using $\hat{\epsilon_t}=\phi(z_t, t, M_F, P)$ . We use Geometric Vector Perceptrons (GVP ) to parameterize $\phi$ because they outperform equivariant neural networks. \citet{satorras2021n}
Following DiffLinker \citet{igashov2022equivariant, jing2020learning, torge2023diffhopp}, we define the "anchor point" as the scaffold atom bonded to the fragment $F$. We ensure the GNN is translationally invariant by first centering the data around the anchor point $a$ and then sampling from $\mathcal{N}(0,I)$ instead of sampling the initial noise from $\mathcal{N}(f(a),I)$ where $f(a)$ is the anchor point center of mass.

During training, we only add noise to coordinates $r$ and feature vector $h$ of the fragment $F$. We keep the scaffold molecule $M_F$ and the protein pocket intact. The input to the neural network is the noised version of fragment $z_t ^{F}$ at time $t$ and the context $u$, which contains the molecular scaffold $M_F$, the protein pocket $P$, and the anchor point $a$. The predicted noise $\hat{\epsilon}^{F_i}$ for the fragment $F_i$ includes coordinates and feature vector $\hat{\epsilon}^{F_i} =[\hat{\epsilon}^{F_i} _x , \hat{\epsilon}^{F_i} _h]$. We only use the predicted coordinates and feature vectors for the fragment atoms and discard the rest.

\cite{hoogeboom2021autoregressive} et al. derived an objective for order agnostic diffusion models to be optimized one step at a time (see \hyperref[sec:ardm]{SI}). Following the ARDM approach, we first sample a random ordering $\sigma$ from the set of all fragment-wise molecule generation permutations $S_D$ at each training step, where $D$ is the number of fragments in the molecule. Next, we uniformly sample a single fragment $F$ to reconstruct with the diffusion model. As proposed by \citet{kingma2021variational}, we use a simplified objective $L(t)=||\epsilon - \hat{\epsilon}_t||^2$ that can be optimized by mini-batch gradient descent.

Additionally, we train a separate model for anchor point prediction (see SI). During sampling, this AnchorGNN (see \hyperref[sec:anchorGNN]{SI}) model predicts the anchor point from among the scaffold atoms. We sample the fragment size from the data distribution conditioned on the pocket size near the anchor point. We repeat the fragment generation process until we reach a maximum number of fragments or molecule size. Algorithms 1 and 2 in \hyperref[sec:alg]{SI} define AutoFragDiff training and sampling procedures, respectively. We compute Lennard-Jones-motivated interactions between generated fragment and pocket atoms to minimize clashes. Following a classifier-based guidance strategy \cite{ho2022classifier}, we compute the gradient of a score function (Lennard-Jones interaction between pocket and fragment) with respect to fragment atom coordinates and add it to them with a negative sign (see \hyperref[sec: steric]{SI}).

\begin{figure*}[h!]
  \includegraphics[scale=0.43]{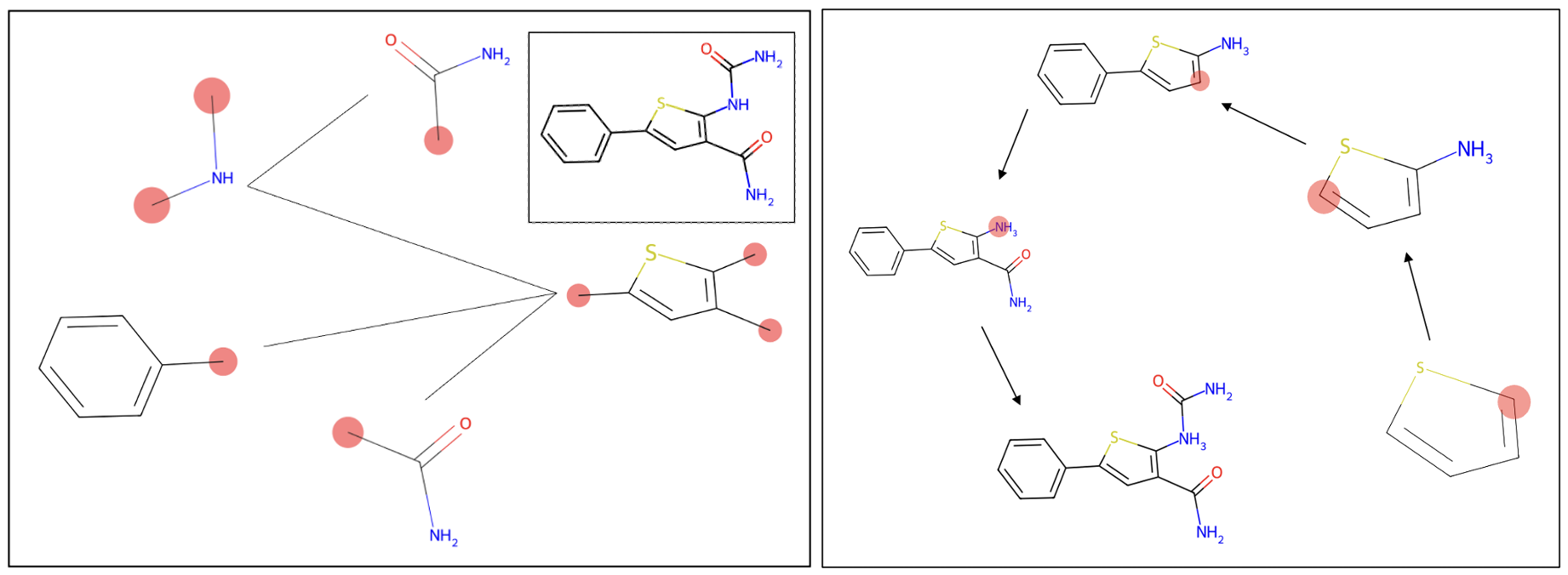}
  \caption{ (\textbf{Left}) The fragment connectivity for a molecule. Highlighted atoms are the anchor points on each fragment. (\textbf{Right}) Sampled generation order for the molecule on the left from its fragments.}
\end{figure*}
\vspace{-0.2cm}

\vspace{-0.2cm}
\section{Datasets}
\vspace{-0.2cm}
\textbf{CrossDock}: We use CrossDock2020\citet{francoeur2020three} to evaluate AutoFragDiff for pocket-based molecule generation. Similar to other studies, we refined the original 22.5 million docked protein binding complexes by filtering for low (<1{\AA}) RMSD and sequence identity of less than 30\%. This procedure yielded 100,000 training complexes and 100 previously unseen testing pockets. We used RDKit \citet{landrum2013rdkit} and BRICS \citet{degen2008art} to fragment molecules by breaking bonds between rings without breaking fused ring systems. We used a maximum of 8 fragments per molecule. We used breadth-first and depth-first search traversals of each molecule's fragment connectivity graph to avoid computing an intractable enumeration of all potential fragment-wise molecule reconstructions. At each reconstruction step, we saved the scaffold atoms and coordinates, the added fragment, and the anchor point where the scaffold connects to the next fragment. Figure 2 illustrates the molecule fragmentation strategy and generation order.

\vspace{-0.2cm}
\section{Results}
\vspace{-0.2cm}

As in TargetDiff\citet{guan20233d}, we use openbabel \cite{o2011open} to reconstruct the molecules from the generated atomic point clouds. In terms of the Jensen Shannon Divergence \citet{lin1991divergence} (JSD) of angles and dihedrals for common ring structures in CrossDock, AutoFragDiff significantly surpasses other models (Table 1). Although it was not a focus of this study, we also assess the generated molecules for various chemical properties (Table 2), including drug-likeliness (QED) and average synthetic accessibility (SA)\citet{ertl2009estimation}. "Diversity" evaluates the average molecular fingerprint similarity across all generated molecule pairs. AutoFragDiff generates realistic molecules with higher calculated binding affinity than the molecules in the test set and exhibits results on par with state-of-the-art models. Figure 8 and Figure 9 (see \hyperref[fig:example]{SI}) show generated molecules for two examples, protein L3MBTL1 (pdb: 2pqw) and P21-activated kinase (pdb: 5i0b).

Additionally, we used PoseCheck \cite{harris2023benchmarking} to evaluate the generated molecules for clashes with protein atoms, strain energies, and interactions with pocket atoms (see \hyperref[sec: add]{SI}). AutoFragDiff molecules averaged 6.7 clashes with pocket atoms (Figure 4), outperforming other diffusion-based models (TargetDiff 9.2 and DiffSBDD 11.8 averages). Non-diffusion models Pocket2Mol and 3DSBDD averaged 5.7 and 3.9 clashes per molecule, while the CrossDock ground truth test set averaged 4.8 clashes. Similarly, non-diffusion modelsPocket2Mol and 3DSBDD generally generated molecules with lower strain energies than diffusion-based models (Figure 5). Considering interaction types, TargetDiff molecules had the most H-bond donors and acceptors (Figure 6), while both AutoFragDiff and TargetDiff showed the most hydrophobic and Van der Waals interactions, on par with the CrossDock test set molecules.  

\begin{table}[h!]
    \centering
    \caption{JSD of angles and dihedrals for most common rings in CrossDock dataset. Best score highlighted in dark gray; second best in light gray. DiffSBDD results are from the conditional all atoms model \cite{schneuing2022structure}}
    \resizebox{\textwidth}{!}{%
    \begin{tabular}{l|cc|cc|cc|cc}
        & \multicolumn{2}{c}{\includegraphics[width=0.1\textwidth]{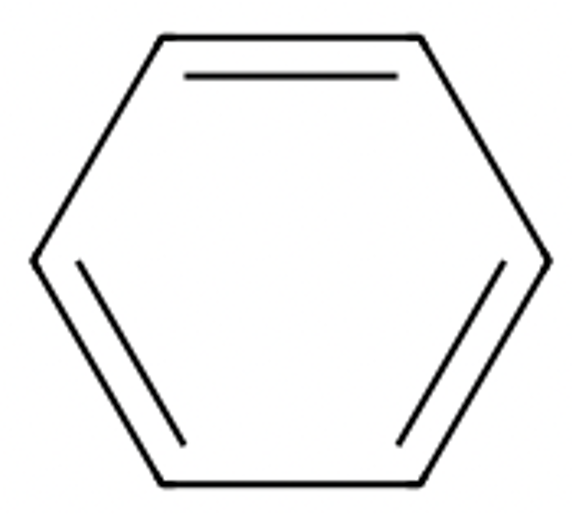}} & 
        \multicolumn{2}{c}{\includegraphics[width=0.1\textwidth]{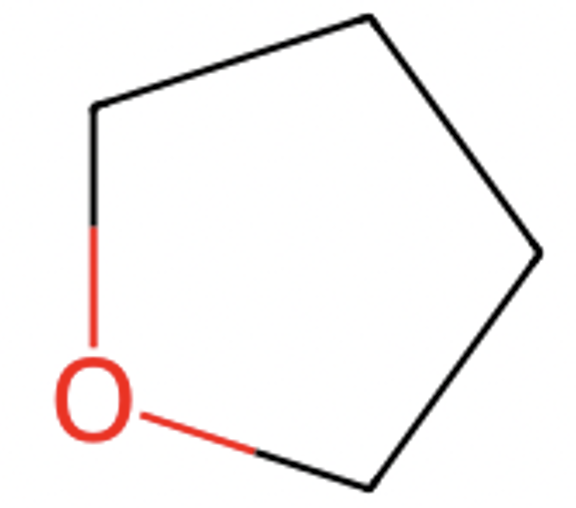}} & 
        \multicolumn{2}{c}{\includegraphics[width=0.1\textwidth]{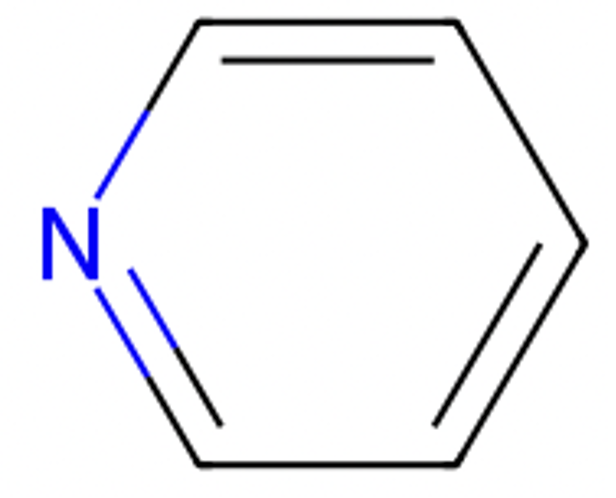}} & 
        \multicolumn{2}{c}{\includegraphics[width=0.11\textwidth]{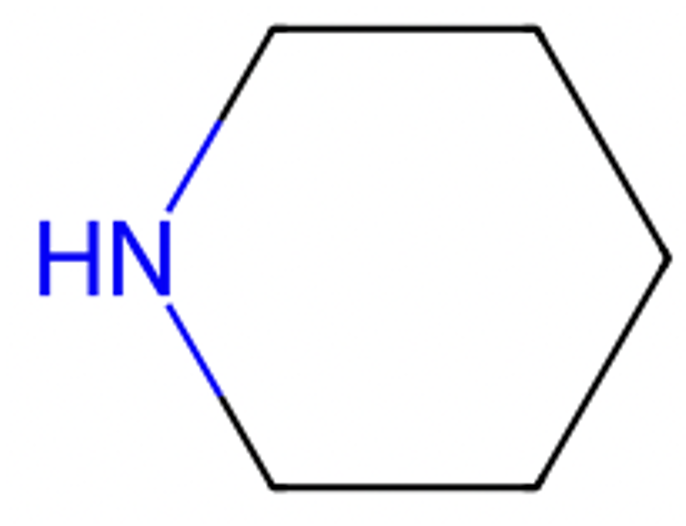}} \\
        \hline
       \textbf{Model} & \textbf{angles} & \textbf{dihedrals} & \textbf{angles} & \textbf{dihedrals} & \textbf{angles} & \textbf{dihedrals} & \textbf{angles} & \textbf{dihedrals} \\
        \cline{1-9}
        3D-SBDD & 0.458& 0.666& 0.293& 0.300& 0.457& 0.625& 0.342& 0.439\\
        Pocket2Mol & 0.438& 0.574& 0.321& 0.272& 0.347& 0.551& 0.408& 0.478\\
        DiffSBDD* & 0.342& 0.549& 0.310& 0.235& 0.254& 0.546& 0.363& 0.435\\
        TargetDiff & \cellcolor{lightgray} \textbf{0.203} & \cellcolor{lightgray} \textbf{0.459} & \cellcolor{darkgray} \textbf{0.154} & \cellcolor{lightgray} \textbf{0.176} & \cellcolor{lightgray} \textbf{0.140} & \cellcolor{lightgray} \textbf{0.460} & \cellcolor{lightgray} \textbf{0.335} & \cellcolor{lightgray} \textbf{0.437}\\
        AutoFragDiff(ours) & \cellcolor{darkgray} \textbf{0.103}& \cellcolor{darkgray} \textbf{0.151} & \cellcolor{lightgray}\textbf{0.191} & \cellcolor{darkgray} \textbf{0.172}&  \cellcolor{darkgray} \textbf{0.073}& \cellcolor{darkgray} \textbf{0.179} & \cellcolor{darkgray} \textbf{0.293}& \cellcolor{darkgray} \textbf{0.333}\\
    \end{tabular}
    }
\end{table}

\begin{table}[h!]
\centering
\caption{Pocket-based generative models comparison. Best score highlighted in dark gray; second best in light gray.}
\begin{tabular}{lcccc}
\toprule
\textbf{Method}       & \textbf{Vina ($\downarrow$)} & \textbf{Diversity ($\uparrow$)} & \textbf{QED ($\uparrow$)} & \textbf{SA ($\uparrow$)} \\
\midrule
3D-SBDD           & -6.71& 0.70& \cellcolor{lightgray}\textbf{ 0.49}& 0.62\\
Pocket2Mol   & -7.15& 0.69& \cellcolor{darkgray}\textbf{0.56} & \cellcolor{darkgray}\textbf{0.74} \\
DiffSBDD& -6.90& \cellcolor{darkgray}\textbf{0.73}& 0.48 &  \cellcolor{lightgray}\textbf{0.63}\\
TargetDiff& \cellcolor{darkgray} \textbf{-7.55} & \cellcolor{lightgray} \textbf{0.72} & 0.49& 0.61\\
AutoFragDiff(ours) & \cellcolor{lightgray} \textbf{-7.45} & 0.69 & 0.45& 0.62\\
CrossDock Test Set Molecules & -7.10& - & 0.47& 0.73 \\
\bottomrule
\end{tabular}
\end{table}

\textbf{Scaffold Extension}: Since our model adds fragments to an existing molecular scaffold at each step, it can further optimize a user-provided starting scaffold. To test the concept, we extracted the Murcko scaffold from every molecule in the CrossDock test set. We augmented each scaffold with up to 4 fragments, generating 20 distinct molecules per CrossDock molecule. 70\% of the newly generated molecules exhibited higher calculated binding affinity than their corresponding starting molecule (average Vina score of -7.8 generated versus -7.1 CrossDock). Figure 3 contains representative examples for an \textit{S. cerevisiae} Cytochrome-\textit{c} peroxidase pocket (pdb: 1a2g).

\begin{figure*}[h!]
  \includegraphics[scale=0.45]{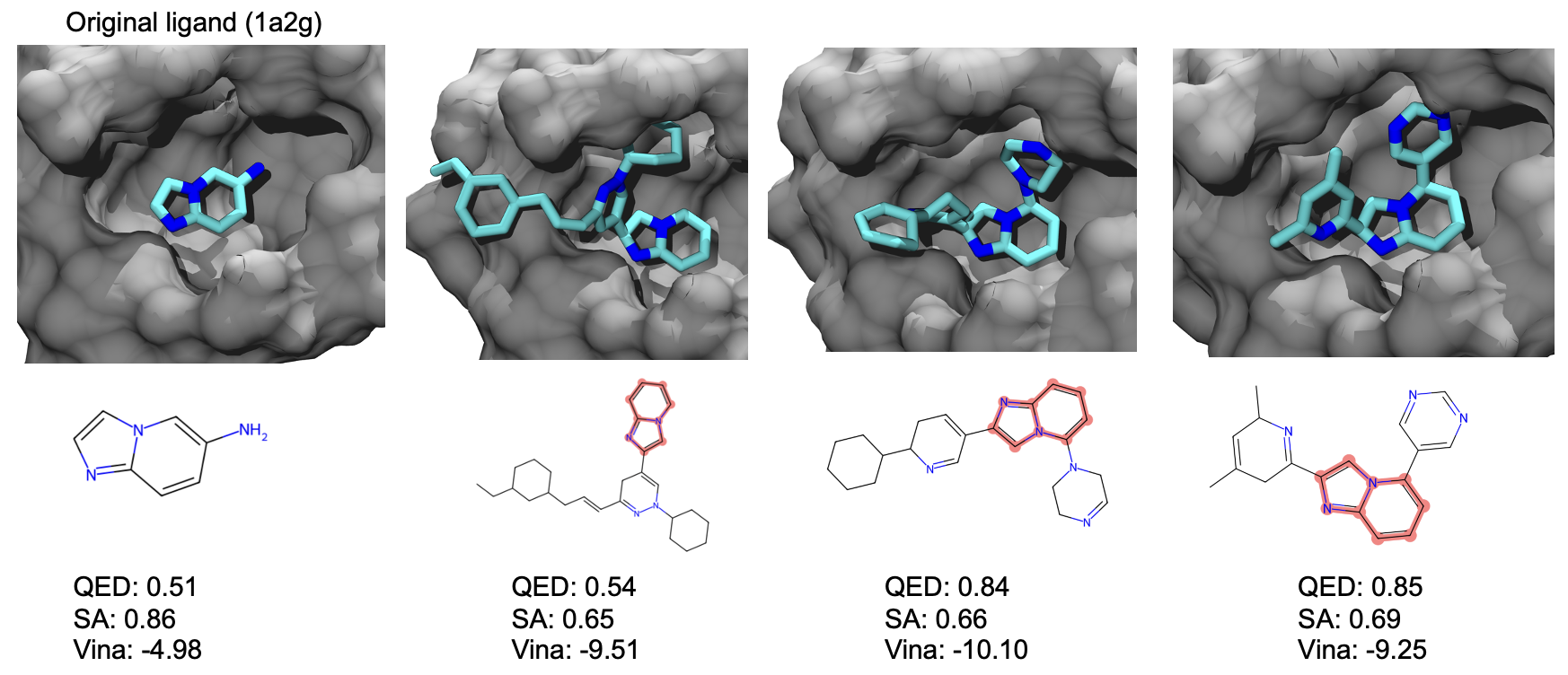}
  \caption{Scaffold (red) extension examples on a Cytochrome-\textit{c} peroxidase (pdb: 1a2g).}
\end{figure*}
\vspace{-0.2cm}
\section{Conclusion}
We introduce AutoFragDiff (\href{https://github.com/keiserlab/autofragdiff}{https://github.com/keiserlab/autofragdiff}), an open-source autoregressive fragment-based diffusion model tailored for pocket-free and pocket-based molecule generation. A standout feature of AutoFragDiff is its capability for scaffold extension, which is a key aspect of many real-world drug design applications, especially in close-in optimization around lead series. The model is adept at generating molecules with high-quality local geometry and exhibits robust binding affinity to target proteins. Looking forward, we aim to enhance the ligand affinity within the pocket using guidance strategies and model architecture improvements.
\bibliography{references}

\newpage

\section{Appendix}

\section*{A.1 Training and Sampling} %
\label{sec: alg}
\begin{algorithm}
\caption{Training}
\label{alg:training}
\begin{algorithmic}[1]
\State \textbf{Input:} Fragment $x^F$, Scaffold $M_F$, anchor point $a$, protein pocket $P$, neural network $\phi$ 
\State Sample Permutation order $\sigma \sim S_D$
\State Sample fragment $F$
\State Sample $t \sim \mathcal{U}(0, \ldots, T)$, \quad $\epsilon_t \sim \mathcal{N}(0, I)$
\State $z^F _t \gets \alpha_t x^F + \sigma_t \epsilon_t$
\State $\hat{\epsilon}_t \gets \phi(z_t, M_F, a, t,$P$)$
\State Minimize $\| \epsilon - \hat{\epsilon}_t \|_2$
\end{algorithmic}
\end{algorithm}

\begin{algorithm}
\caption{Sampling}
\label{alg:sampling}
\begin{algorithmic}[1]

\For{$i$ in $1..D;$}
\State \textbf{Input:} Scaffold $M_{F_i}$, anchor point $a_i$, protein pocket $P$, neural network $\phi$
\State Center everything at $f(a_i)$
\State Sample $z_T ^{F_i} \sim \mathcal{N}(0, I)$
\For{$t$ in $T; T - 1; \ldots ; 1$}
    \State Sample $\epsilon_t \sim \mathcal{N}(0, I)$
    \State $\hat{\epsilon}_t \gets \mathcal \phi(z^{F_i} _t, t, M_{F_i}, a_i, P)$
    \State $z_{t-1} ^{F_i} \gets (1/ \bar{\alpha}_t) \cdot z_t - \bar{\sigma}_t ^2 / (\bar{\alpha}_t \sigma_t) \cdot \hat{\epsilon}_t + \zeta_t . \epsilon$ 
\EndFor
\State Sample $x^{F_i} \sim p(x^{F_i}|z_0 ^{F_i}, M_{F_i}, a_i, P)$
\EndFor
\end{algorithmic}
\end{algorithm}

For sampling molecule sizes, we first bin the pocket volumes into 10 bins (using grids inside the protein pocket) and find the distribution of molecule sizes for each bin. During sampling, we sample molecule sizes from the distribution of the corresponding volume bin. For the first generation step, the anchor point is selected from the pocket atoms in contact with the original ligand. We first bin the pocket volume within 3.5 \text{\AA} of the anchor point for fragment size and then sample the fragment sizes from the corresponding bin. The average size of generated molecules from our model is 26 atoms.

\section*{A.2 Diffusion Process}

At each timestep $t=0...T$ the conditional distribution of the intermediate state $z_t ^F$  for a single fragment $F$ given the previous state is defined by the multivariate normal distribution: 

\begin{equation}
q(z_t^F | z_{t-1}^F) = N\left(z_t^F; \bar{\alpha}_t z_{t-1}^F, \bar{\sigma}_t^2 I\right)
\end{equation}

In this equation $\bar{\alpha}_t = \alpha_t / \alpha_{t-1}$ controls how much signal is retained and $ \bar{\sigma}_t = \sigma^2 _t - \bar{\alpha}^2 _t \sigma ^2 _{t-1}$ controls how much noise is added. The full transition model for diffusion is Markovian:

\begin{equation}
q(z_0 ^F, z_1 ^F, ..., z_T ^F | x^F) = q(z_0 ^F | x^F) \prod_{t=1}^{T} q(z_t ^F | z_{t-1} ^F)
\end{equation}

The true denoising process has a closed-form solution when conditioned on $x^F$: 
\begin{equation}
q(z_{t-1} ^F | z_t ^F, x^F) = N\left(z_{t-1} ^F; \mu (x, z_t), \zeta_t ^2 I \right)
\end{equation}

where $\mu_t (x^F, z_t ^F)$ and $\zeta_t$ have analytical solutions:

\begin{equation}
    \mu_t (x^F, z_t ^F) = \frac{\bar{\alpha}_t \sigma ^2 _{t-1}}{\sigma ^2 _t} z_t + \frac{\alpha_s \bar{\sigma}^2 _t}{\sigma ^2 _t},   \quad \zeta_t = \frac{\bar{\sigma _t} \sigma _{t-1}}{ \sigma_t}
\end{equation}

We trained AutoFragDiff with $T=500$ diffusion steps using a polynomial noise scheduler:

\begin{equation}
    \alpha_t = (1-2s) \cdot (1 - (t/T)^2)
\end{equation}

where $s=10^{-5}$ is the precision value to help with numerical issues.

\section*{A.3 Geometric Vector Perceptrons}
\label{sec: GVP}
GVP \citet{jing2020learning} uses nodes with scalar features $s$ as inputs. These scalars represent embedded features of atoms without accompanying vector features. Edges within the graph incorporate a normed direction vector alongside the distance between two nodes. More specifics about this can be found in GVP paper.\citet{jing2020learning} 

As described previously in DiffHopp \citet{torge2023diffhopp}, the attributes of nodes and edges undergo linear transformations. Edge embeddings are achieved in two phases: initially, their inputs are normalized using layer normalization \citet{ba2016layer}, and following this, they are channeled through a GVP. Here, both $\sigma$ and $\sigma^+$ operate as the identity function, resulting in a scalar with a hidden size of $h/2$ and a singular vector. Nodes undergo a parallel embedding process, culminating in outputs of $h$ scalars and $h/2$ vectors, summing up to $h$ features. The message-passing layers can be expressed as:

\begin{equation}
\begin{aligned}
    m_{vw} ' &= \phi_e (h_v, h_w, e_{vw}), \\
    m_v ' &= \sum_{w \in N_v} \tilde{e}_{vw} m_{vw} \\
    h_v ' &= \phi_h(h_v, m_v ')
\end{aligned}
\end{equation}

Within this equation, $\tilde{e}_{vw} = \phi_{att}(m_{vw})$ acts as an attention mechanism, enabling the learning of soft edge estimates, mirroring the approach in EGNN. The function $\phi_e$ combines three GVPs featuring hidden sizes $(h, h/2)$. Notably, the final GVP has $\sigma$ as its identity function. Meanwhile, $\phi_{att}$ embodies a single GVP translating to a singular scalar with $\sigma$ functioning as the sigmoid activation. A factor of $C=100$ normalizes the resulting output.

The relationship between $\phi_h(h_v, m_v ')$ is captured by the equation $\phi_h(h_v, m_v ') = \text{norm}(h_v + \phi_h ' (\text{norm}(h_v + m_v ')))$. This employs a residual architecture where $\phi_h ' $ integrates two GVPs with sizes $(h, h/2)$. This encapsulates input, hidden, and output dimensions. The terminal layer once again adopts $\sigma$ as the identity function. The term "norm" represents layer normalization, which isn't applied to vectors.

\section*{A.4 Autoregressive Diffusion Models}
\label{sec:ardm}
Autoregressive models can factorize a multivariate distribution into a product of $D$ univariate distributions. 

\begin{equation}
\log p(x) = \sum_{t=1}^{D} \log p(x_t|x_1, ..., x_{t-1})
\end{equation}

Sampling from such models can be done through $D$ iterative sampling steps. Order agnostic models can generate variables with random orderings $\sigma \in S_D$ where $S_D$ is the set of all permutations for building the molecule from its fragments. The log-likelihood of these models can be written as: 

\begin{equation}
\log p(M|P) \geq \mathbb{E} _{\sigma \sim \mathcal{U}(S_D)} \sum_{i=1} ^{D} \log p(M_{\sigma _i} | M_{\sigma (<i)}, P) 
\end{equation}

In this equation, $M$ is the set of molecule atoms, $P$ is the set of protein atoms, and $M_{\sigma _i}$ is the molecule generated with sampled ordering $\sigma$ at the fragment step $i$. Hoogeboom et. al. derived an objective for order agnostic diffusion models \citet{hoogeboom2021autoregressive} that only needs to be optimized for a single step at a time:

\begin{equation}
\log p(M|P) \geq \mathbb{E}_{\sigma \sim U(S_D)} D. \mathbb{E}_{i \sim \mathcal{U}(1..D)}  \log p(M_{\sigma _i} | M_{\sigma (<i)}, P) 
\end{equation}

According to this objective, during training, we sample a random order $\sigma$ of molecule generation uniformly from the set of all generation orders $S_D$, and a single fragment from the uniform distribution of all fragments in the molecule. We train the diffusion model to predict this single fragment. In practice, we optimize a simplified $L(t)=||\epsilon - \hat{\epsilon}_t||^2$ loss by mini-batch gradient descent.  

\section*{A.5 Hyperparameters}
\label{sec: hyp}

We consider the protein graph as the protein atoms within 7 \text{\AA} of the original ligand. Edges within the ligand are fully connected, while protein-ligand and protein-protein edges are drawn with a radius threshold of 4.5 \AA. The edge features for nodes $i$ and $j$ are the distance $d_{ij}$ and the normalized direction vector $(x_i - x_j)/d_{ij}$.  As previously suggested by \citet{hoogeboom2022equivariant}, we scale node types $h$ by a factor of 0.25. The final model had 6 GVP layers, with hidden dimension of 128 and a joint embedding dimension of 32. We trained the model with a learning rate of $2\times 10^{-4}$ for 500 epochs.

\section*{A.6 Additional results}
\label{sec: add}

Table 3 compares different models by JSD for different types of bonds.

\begin{table}[h!]
\caption{Comparison of JSD of bond distances in different bond types for different models. Best results per row are highlighted in dark gray, and second best in light gray.}
\small
\centering
\begin{tabular}{c|c|c|c|c|c} 
\midrule
\textbf{Bond} & \textbf{3D-SBDD} & \textbf{Pocket2Mol} & \textbf{DiffSBDD}&  \textbf{TargetDiff}& \textbf{AutoFragDiff} \\ 
\midrule
\textbf{C-C} & 0.576& 0.455&  \cellcolor{lightgray} 0.347& \cellcolor{darkgray}\textbf{0.286} &   \textbf{0.363}\\
\textbf{C=C}  & 0.421& 0.561& 0.314& \cellcolor{darkgray} \textbf{0.220} & \cellcolor{lightgray}  \textbf{0.221} \\
\textbf{C-N}  & 0.383& 0.321& 0.313& \cellcolor{darkgray} \textbf{0.242 }& \cellcolor{lightgray} \textbf{0.290} \\
\textbf{C=N}  & 0.443& 0.377& 0.348& \cellcolor{darkgray} \textbf{0.179} & \cellcolor{lightgray} \textbf{0.231} \\
\textbf{C-O}  & 0.394& \cellcolor{lightgray} 0.326& 0.353& \cellcolor{darkgray} \textbf{0.298} &  \textbf{0.354} \\
\textbf{C=O}  & 0.511& 0.446& 0.398& \cellcolor{lightgray} \textbf{ 0.398} &  \cellcolor{darkgray} \textbf{0.363} \\
\textbf{C:C}  & 0.459& 0.309& 0.316& \cellcolor{darkgray} \textbf{0.176} & \cellcolor{lightgray}  \textbf{0.295} \\
\textbf{C:N}  & 0.582& 0.377& 0.348& \cellcolor{darkgray} \textbf{0.158} & \cellcolor{lightgray}  \textbf{0.217} \\
\bottomrule
\end{tabular}
\end{table}

As described below, we used PoseCheck \cite{harris2023benchmarking} to evaluate each model's generated molecules for their number of clashes with pocket atoms, strain energies, and interactions (hydrogen bond donors and acceptors, hydrophobic interactions, and Van der Waals contacts) with pocket atoms.

\subsection{Steric clashes}
\label{sec: steric}
We computed steric clashes of generated molecules with protein atoms with a clash tolerance of 0.5 \AA, as described in PoseCheck. We use a classifier guidance approach to minimize the clashes to pocket atoms in our model. We calculate the Lennard-Jones (LJ) interaction of the fragment atoms with pocket atoms as the guidance function and add the negative gradient of this score with respect to fragment coordinates to the current fragment atom coordinates. When computing the Lennard-Jones interactions, we include pocket hydrogen atoms, although the ligand diffusion itself does not include explicit hydrogens.

\begin{equation}
U(r) = 4 \varepsilon \left[ \left( \frac{\sigma}{r} \right)^{12} - \left( \frac{\sigma}{r} \right)^{6} \right]
\end{equation}

In this equation, $\sigma$ is the sum of Van der Waals radii of the pocket and ligand atoms, and $r$ is the distance between protein and ligand atoms. We clip the output at 1000 to avoid very large values. 

\begin{equation}
    x^F = x^F - \epsilon \nabla _{x^F} U(r)
\end{equation}

We use a cosine-$\beta$ weight scheduler to progressively lower the effect of LJ guidance over the inference trajectory. Note that the LJ guidance is only used for avoiding clashes with pocket atoms and does not have a physical meaning. CrossDock test set molecules have on average 4.8 clashes with pocket atoms. Our model shows an average of 6.7 clashes, outperforming other diffusion models Targetdiff: 9.2 and DiffSBDD: 11.8. However, non-diffusion-based models have fewer clashes with pocket atoms, with Pocket2Mol having an average of 5.7 and 3DSBDD 3.9 clashes. 

\begin{figure*}[h!]
    \centering
  \includegraphics[width=1\linewidth]{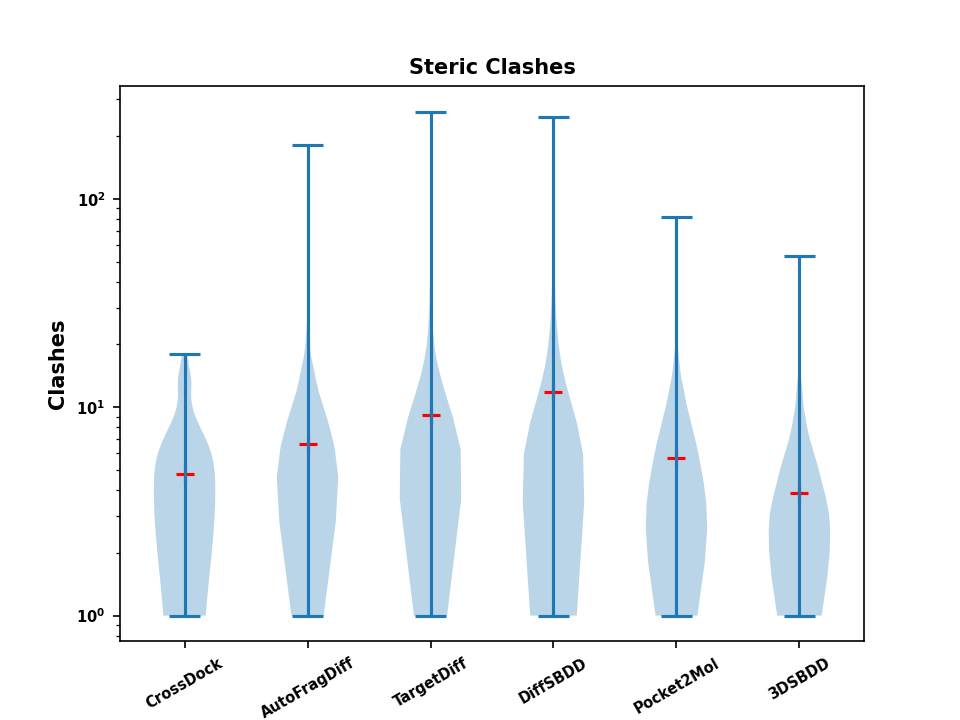}
  \caption{Steric clashes of different models.}
\end{figure*}

\subsection{Strain energy}
We computed molecule strain energies as the difference between the internal energy of the generated and minimized conformers, using the Universal Force Field (UFF) from RDKit (Figure 5). Diffusion-based models all showed higher strain energies than non-diffusion models with Pocke2Mol model having the lowest strain energy.

\begin{figure*}[h!]
    \centering
  \includegraphics[scale=0.7]{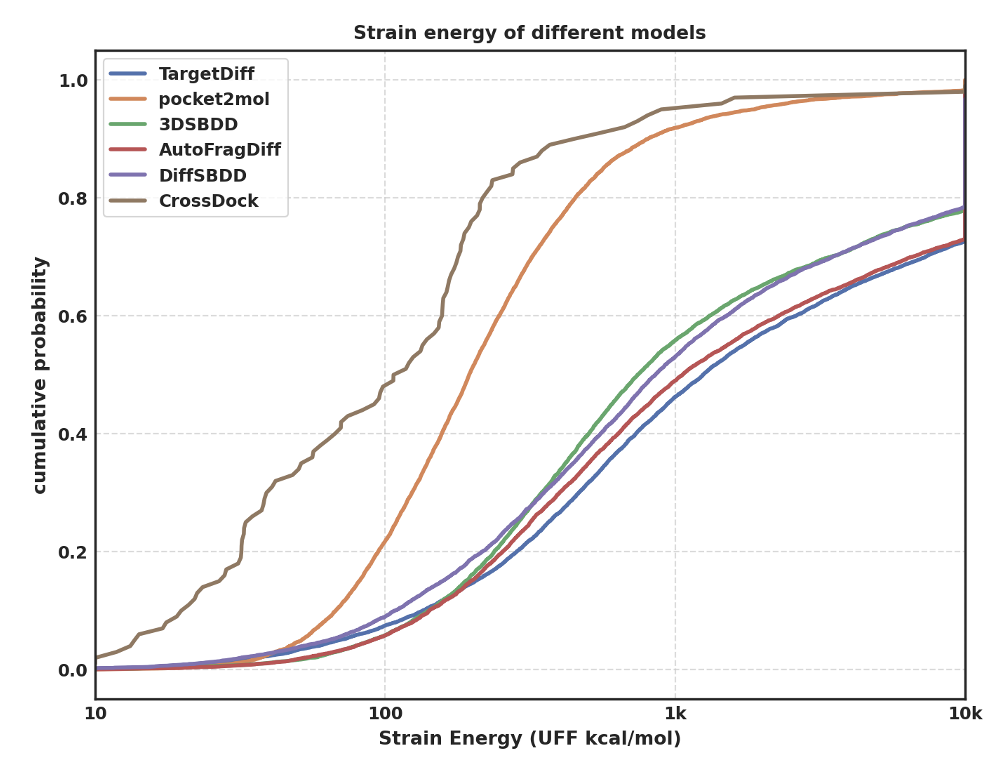}
  \caption{Strain energies of different models}
\end{figure*}

\subsection{Ligand-pocket interactions}
Using PoseCheck, we computed four interaction types between molecule poses and pocket atoms (hydrogen bond donors, hydrogen bond acceptors, hydrophobic interactions, and Van der Waals contacts) (Figure 6). CrossDock molecules have more hydrogen bond donors than any generated molecules. TargetDiff shows more H-bond donors and acceptors than the other generative models. In terms of hydrophobic and Van der Waals interactions, AutoFragDiff and TargetDiff show similar performance which is also on par with CrossDock test set molecules.

\begin{figure*}[h!]
    \centering
  \includegraphics[width=1\linewidth]{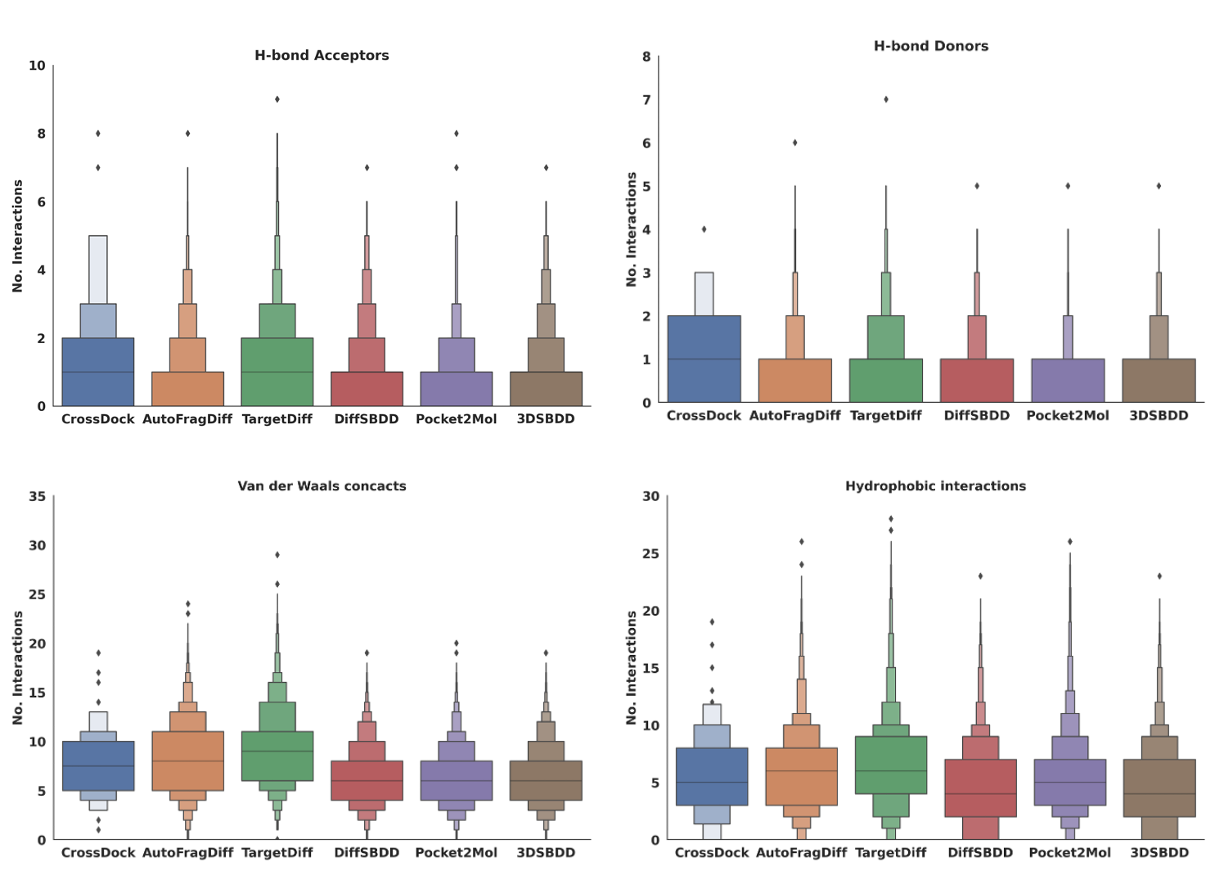}
  \caption{Different interaction types of generated poses with protein atoms.}
\end{figure*}

\section*{A.7 Fragmentation}
\label{sec: frag}

 During fragmentation, we first break all the bonds between rings without breaking fused ring systems. In addition, we also use RDKit and use BRICS to fragment the molecules (Figure 7). A maximum of 8 fragments is used for each molecule; if the number of fragments exceeds 8, we connect the smallest fragments iteratively until the maximum of 8 fragments is reached. Given the fragment connectivity of a molecule (fragment adjacency), we compute all BFS and DFS traversals of the molecule graph based on the fragment. We only use BFS And DFS traversals to avoid computing all of the molecule's fragmentation graph's traversals for computational feasibility. A generation order defines how fragments are added step-wise based on their connectivity to make a complete molecule. At each step of the generation order, we save the scaffold atoms and their coordinates, the added fragment at the generation step and its coordinates, the generation step, and the scaffold anchor point for the next fragment. This strategy greatly augments the size of the dataset as well.

\begin{figure*}
  \centering
  \includegraphics[scale=0.45]{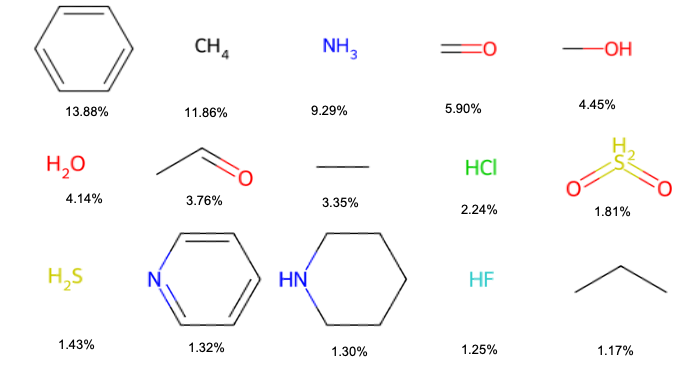}
  \caption{Top 15 occurring fragments using our custom fragmentation in CrossDock dataset.}
\end{figure*}

\section*{A.8 Anchor point predictor}
\label{sec:anchorGNN}
We trained a standalone neural network (AnchorGNN) to predict the anchor points during sampling. We used graph convolutional layers (GCL) to predict the probability of each scaffold atom being an anchor point. The molecular scaffold and protein pocket atoms are the model inputs. We use one-hot encoded atom types as scaffold atom features. For pocket atoms, the features are atom types, amino acid types, and whether the atom belongs to the backbone or sidechain. We use inter-atomic squared distance $d_{ij} ^2=||r_i - r_j||^2$ as the edge feature.

The update for feature $h$ and coordinates of node $i$ at layer $l$ are computed as follows:

\begin{equation}
    m_{ij} = \phi _e (h_i ^l, h_j ^l, d_{ij} ^2), \quad h_i ^{l+1} = \phi_h (h_i ^l, \sum_{j \neq i} {m_{ij}}), \quad r_i ^{l+1} = r_i ^l + \phi_{vel} (r^l, h^l, i) 
\end{equation}

with $d_{ij} = ||r_i - r_j||$ and $\phi_e, \phi_h$ being learning functions. We perform a sequence of $l$ Graph Convolutional Layers ($l=4$). Finally, node embeddings for the scaffold molecule $h^M$ are linearly transformed to a single number and passed through a sigmoid function to compute the probabilities. A binary cross-entropy loss is used to train the model. During sampling, we take the anchor point with the highest probability. A learning rate of $5\times 10^{-4}$ was used to train this model.

\begin{figure*}
  \centering
  \includegraphics[scale=0.5]{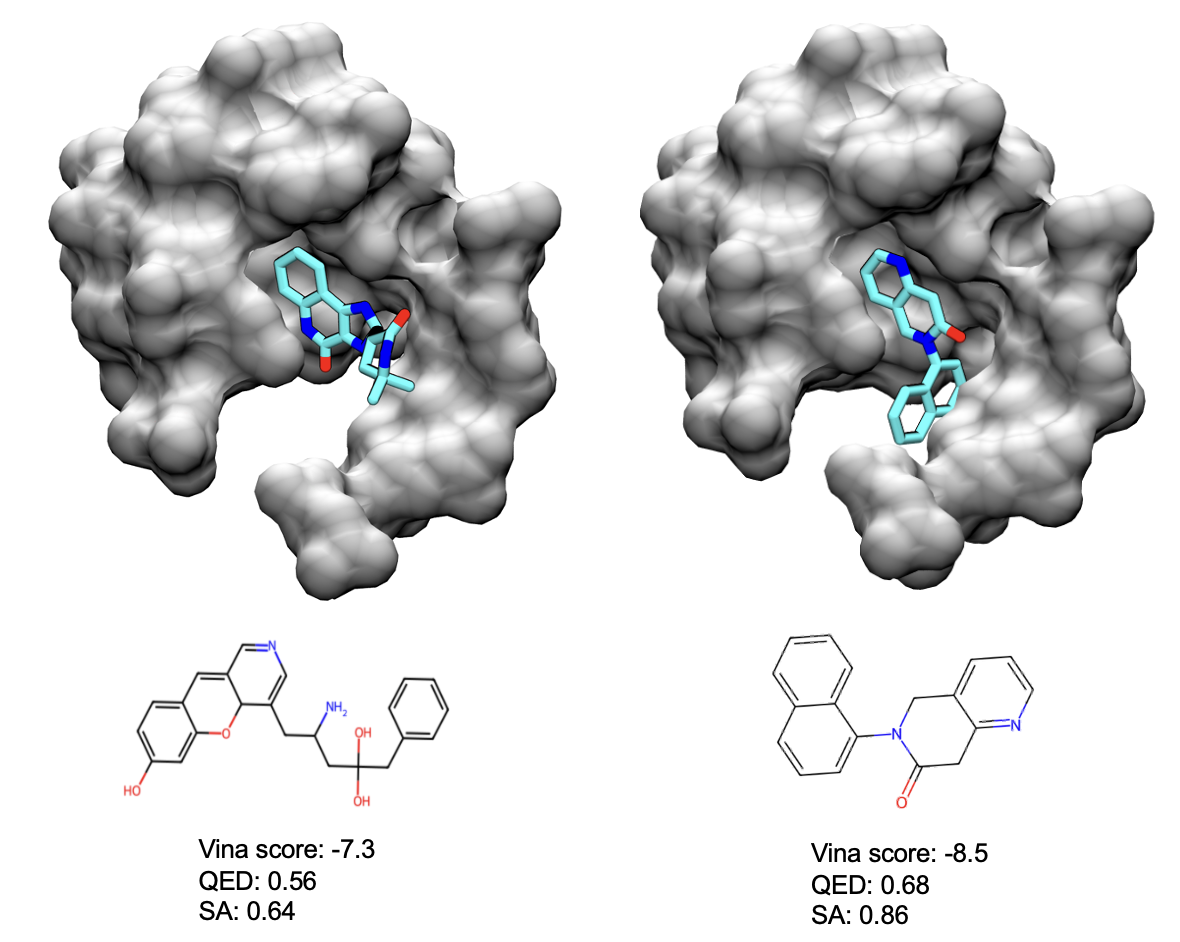}
  \caption{Examples of generated molecules for protein L3MBTL1 (pdb: 2pqw). }
  \label{fig:example}
\end{figure*}

\begin{figure*}
  \centering
  \includegraphics[scale=0.5]{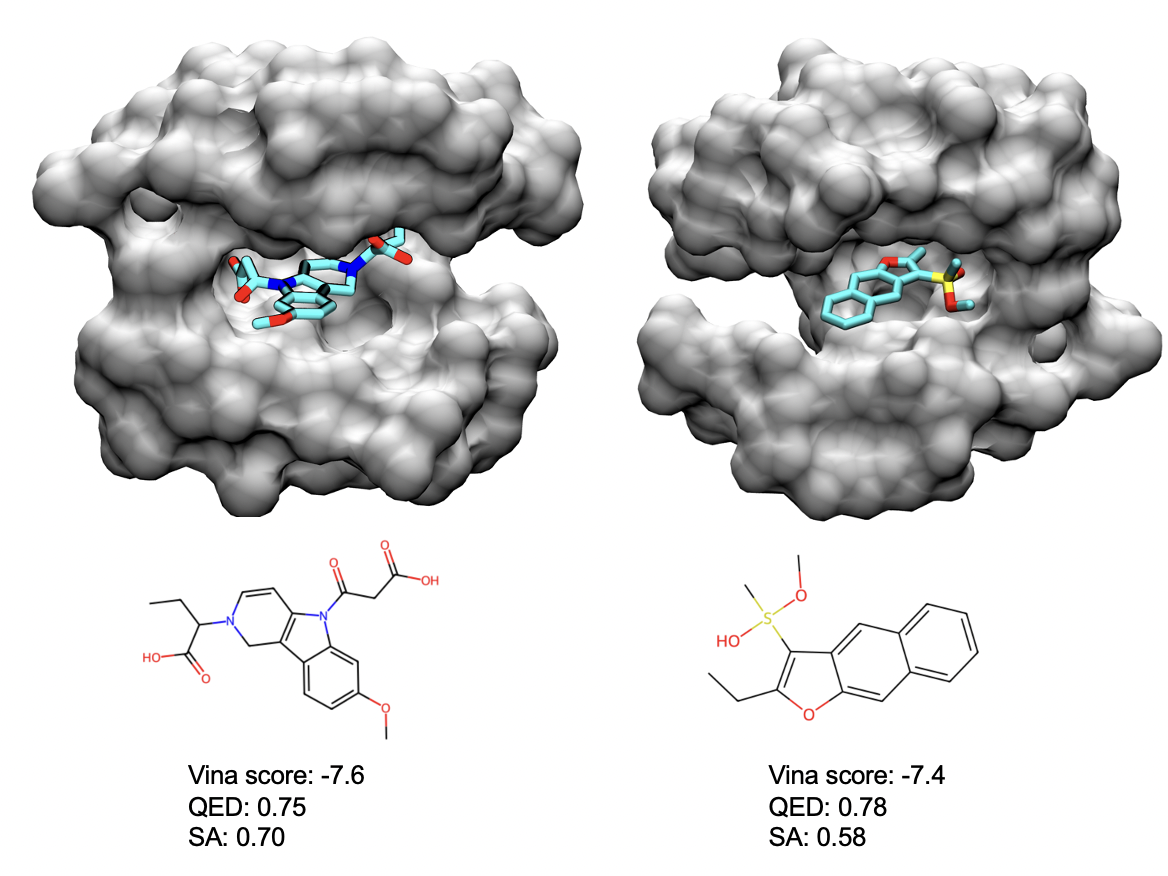}
  \caption{Examples of generated molecules for P21-activated kinase (pdb: 5i0b).}
\end{figure*}

\end{document}